\documentclass[a4paper]{article}

\usepackage{INTERSPEECH2022}

\title{Transsion Multilingual Speech Recognition System for MLC-SLM 2025 Challenge}
\name{Xiaoxiao Li$^1$, An Zhu$^1$, Youhai Jiang$^1$, Fengjie Zhu$^1$}
\address{
  $^1$Shenzhen Transsion Holdings Co., Ltd, China}
\email{\{xiaoxiao.li7, an.zhu, youhai.jiang, fengjie.zhu\}@transsion.com}

\begin{document}

\maketitle
\begin{abstract}
This paper presents the architecture and performance of a novel Multilingual Automatic Speech Recognition (ASR) system developed by the Transsion Speech Team for Track 1 of the MLC-SLM 2025 Challenge. The proposed system comprises three key components: 1) a frozen Whisper-large-v3 based speech encoder, leveraging large-scale pretraining to ensure robust acoustic feature extraction; 2) a trainable adaptor module using Linear-ReLU-Linear transformation mechanisms to effectively align speech and text representations; and 3) a frozen Qwen2.5-7B-Instruct large language model (LLM) integrated with trainable LoRA for optimized contextual linguistic decoding. By systematically combining pretrained models with task-specific fine-tuning, the system achieved a word/character error rate (WER/CER) of 9.83\% across 11 languages in the evaluation set and ranked third place among global participants.

\end{abstract}

\section{Introduction}

Large Language Models (LLMs) have emerged as transformative foundation models across language processing tasks, with growing applications in speech and audio domains including automatic speech recognition (ASR) and spoken dialogue systems\cite{chu2023qwen}\cite{chu2024qwen2}\cite{tang2023salmonn}\cite{zhang2023speechgpt}\cite{ma2024embarrassingly}. However, the scarcity of real-world conversational speech data , especially in multilingual contexts, poses a significant challenge to advancing the field\cite{ardila2019common}\cite{chen2021gigaspeech}\cite{yang2024gigaspeech}\cite{li2024msr}. To overcome this limitation, MLC-SLM 2025 challenge introduces a novel multilingual conversational speech dataset, featuring 1,500 hours of real-world dialogue recordings across 11 diverse languages. This dataset is designed to support the development of multilingual LLM-based ASR models, providing valuable resources to the speech community.

In this paper, we presents the Multilingual Automatic Speech Recognition (ASR) system submitted by the Transsion Speech Team for Track 1. The rest of the paper is structured as follows: Section 2 provides an summary of the datasets utilized for the competition, while Section 3 elaborates on the system architectures, training methodologies, and ablation studies.

\section{DATASET}

The competition provides a multilingual conversational speech corpus consisting of 1,500 hours of real-world conversational speech recordings across 11 diverse languages: English (en), French (fr), German (de), Italian (it), Portuguese (pt), Spanish (es), Japanese (ja), Korean (ko), Russian (ru), Thai (th), and Vietnamese (vi), representing a wide range of linguistic diversity.

In addition, to enhance the generalization capability of the models, we utilized the open-source MSR-86K\cite{li2024msr} dataset as an external resource. However, due to time constraints, we randomly selected a subset of the MSR-86K dataset corresponding to the above 11 languages in our experiment. Detailed duration information about the speech datasets used in this competition is provided in Table 1.

\begin{table}[h]
    \caption{Speech duration summary(hrs)}
    \label{tab:example}
    \centering
    \begin{tabular}{ccc}
    \toprule
        \textbf{ } & \textbf{competition speech\ } & \textbf{\ MSR-86K subset}  \\ \midrule
        English & 500 & 948  \\ 
        French & 100 & 550  \\
        German & 100 & 761  \\
        Italian & 100 & 546  \\
        Portuguese & 100 & 720  \\
        Spanish & 100 & 550  \\
        Japanese & 100 & 496  \\
        Korean & 100 & 427  \\
        Russian & 100 & 550  \\
        Thai & 100 & 784  \\
        Vietnamese & 100 & 584  \\
    \bottomrule
    \end{tabular}
\end{table}

\section{System Description}

The overall multilingual ASR architecture is illustrated in Figure 1. The system is designed based on the standard Encoder-Adaptor-LLM architecture\cite{ma2024embarrassingly}, comprising the following components:\\
\noindent\textbf{Frozen Speech Encoder: }The system utilizes the Whisper-large-v3\cite{radford2023robust} encoder, a pre-trained model capable of processing 128-dimensional Mel filterbank features as input. This encoder generates continuous representations of the input speech at a temporal resolution of 20ms per frame. The frozen nature of the encoder ensures that its robust pre-trained capabilities are retained during the training process.

\noindent\textbf{Trainable Adaptor: }To achieve seamless integration between the speech encoder and the text-based LLM, a trainable adaptor network is employed. At the initial stage, a frame-splicing operation reduces the temporal resolution from 20ms to 40ms per frame. This operation significantly decreases the sequence length, thereby improving the computational efficiency of the downstream LLM. 
After the frame-splicing operation, a Linear-ReLU-Linear neural network is employed to transform the output dimensions of the speech encoder into the input embedding space of the LLM. Specifically:
1)First Linear Layer: The input dimension is determined by the speech encoder's output dimension multiplied by the frame-splicing factor (2 in this case), resulting in an input size of 2560. This layer projects the input to a space matching the LLM's embedding dimension, which is 3584.
2)ReLU Activation: A ReLU activation function is applied after the first linear transformation to introduce non-linearity.
3)Second Linear Layer: This layer takes the output of the ReLU activation as input, with both the input and output dimensions set to the LLM's embedding dimension (3584).

\noindent\textbf{Frozen LLM with Trainable LoRA: }The LLM component is initialized using pre-trained weights from Qwen2.5-7B-Instruct\cite{yang2024qwen2technicalreport}\cite{qwen2025qwen25technicalreport}. To adapt the LLM for ASR task, the system employs Low-Rank Adaptation (LoRA), a parameter-efficient fine-tuning method. This involves training only a small subset of additional parameters while keeping the majority of the LLM's weights frozen. The LoRA configuration in this system uses a rank of 64 and a scaling factor ( \text{lora\_alpha} ) of 16, enabling effective task-specific adaptation without affecting the pre-trained capabilities of the LLM.

\begin{figure}[t]
  \centering
  \includegraphics[width=\linewidth]{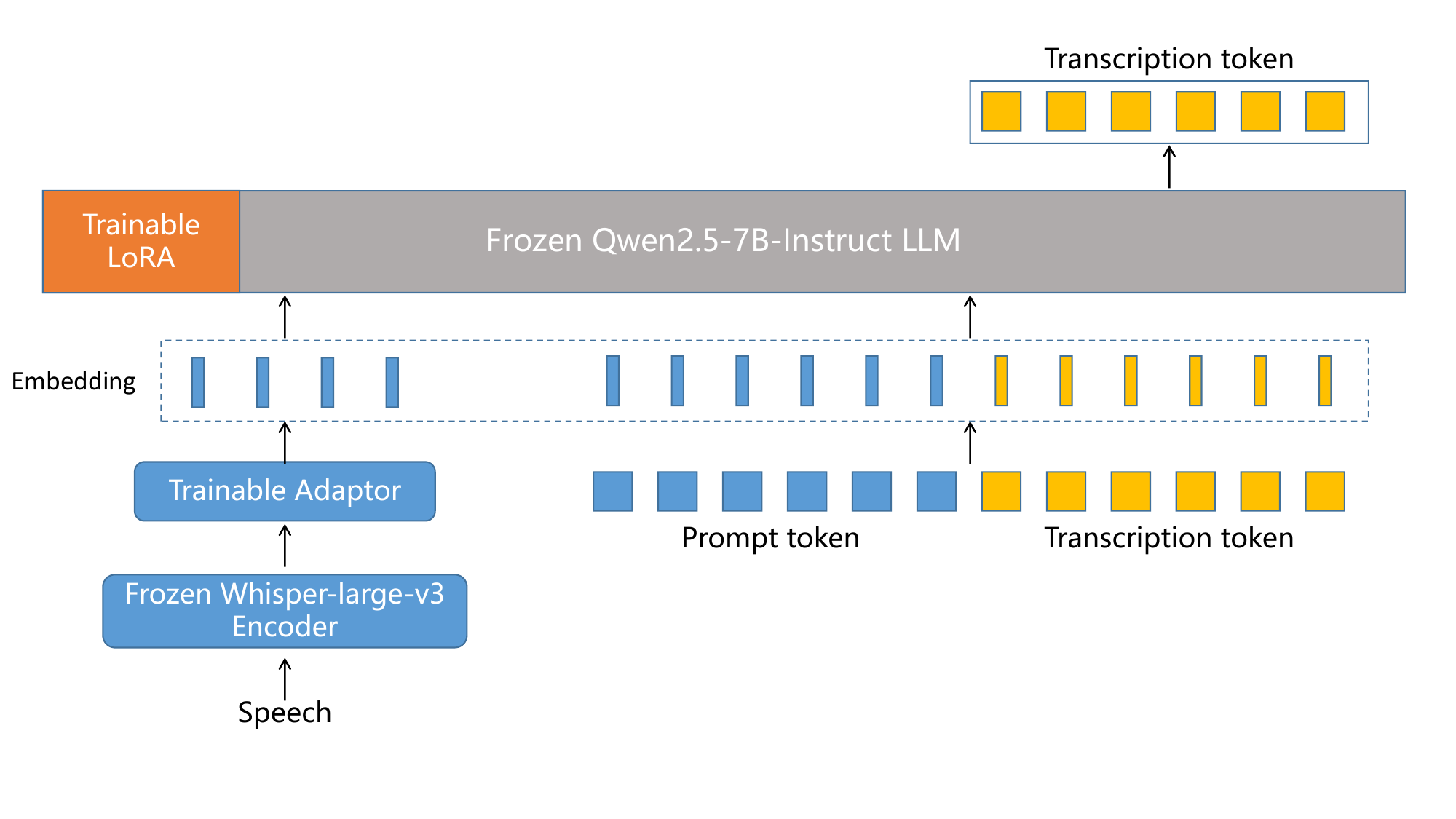}
  \caption{Proposed Multilingual ASR System Architecture.}
  \label{fig:speech_production}
\end{figure}

\subsection{Training Methodologies}
\noindent\textbf{Text Normalization: }During the training process, punctuation was removed, and all text in the training dataset was converted to lowercase, in alignment with the official baseline. Subsequently, the transcriptions were structured following the Qwen chat template, as illustrated below:

\begin{verbatim}
<|im_start|>system
You are a helpful assistant.
<|im_end|>
<|im_start|>user
Transcribe speech to text.
<|im_end|>
<|im_start|>assistant
Transcription
<|im_end|>
\end{verbatim}

\noindent\textbf{Training Parameters: }The training process was conducted using eight NVIDIA A100 GPUs, each equipped with 80GB of GPU memory. A batch size of 4 was assigned per GPU, with gradient accumulation steps configured to achieve an effective total batch size of 64. The optimization utilized the Adam optimizer with an initial learning rate of 1e-4, which was linearly decayed to zero over the course of 1,000,000 steps. A warm-up phase was implemented for the first 1,000 steps to stabilize the learning process. The model was trained for a total of 4 epochs.

\noindent\textbf{Checkpoint Average: }Upon completing the training, model checkpoints were evaluated based on the word/character error rate (WER/CER) across 11 languages in the development(Dev) set. The best-performing checkpoints were selected and subsequently averaged to achieve optimal performance.

\subsection{Ablation Studies}

The ablation study presented in Table 2 highlights the progressive improvements on the development(Dev) set achieved through data augmentation and model scaling. Our baseline system, utilizing competition speech data and the Qwen2-1.5B-Instruct model, achieves a WER/CER of 11.57\%, significantly surpassing the official baseline. Expanding the training dataset by incorporating the MSR-86K subset further reduces the WER/CER to 10.79\%, demonstrating the positive impact of diversified training data.

The best performance, with a WER/CER of 10.01\%, is achieved by simultaneously scaling both model capacity (using Qwen2.5-7B-Instruct) and enhancing training data diversity. This result underscores the synergistic benefits of model scaling and data augmentation strategies. The systematic progression of results confirms that both architectural advancements and improvements in data diversity are critical to enhancing speech recognition performance.

For comparative analysis, we fine-tuned the Whisper models in multilingual configurations, as illustrated in Figure 2. The Whisper Turbo model achieves a WER/CER of 12.97\% using only competition speech data, while Whisper-large-v3 reaches 11.61\% under identical conditions. Augmenting Whisper-large-v3 with the MSR-86K subset reduces the WER/CER to 10.4\%, demonstrating similar data augmentation benefits as observed in Qwen models. This positions Whisper-large-v3 competitively close to our Qwen2-1.5B-Instruct model (10.79\%), though still trailing the best Qwen2.5-7B-Instruct configuration.

Additionally, we explored replacing the encoder with our finetuned Whisper-large-v3 encoder in Qwen2.5-7B-Instruct configuration. However, the initial results did not surpass those of the original setting. We leave further investigation of this approach to future studies.

For all the ablation studies, a decoding beam size of 6 is employed during the inference generation process.

\begin{table}[h]
    \caption{Dev WER/CER results of Ablation Studies.}
    \label{tab:example}
    \centering
    \begin{tabular}{c p{2.5cm} c}
    \toprule
        \textbf{ } & \textbf{Train Dataset} & \textbf{WER/CER} \\ \midrule
        Track 1 official baseline  & competition speech & 21.49\% \\ 
        Qwen2-1.5B-Instruct & competition speech & 11.57\% \\  
        Qwen2-1.5B-Instruct & competition speech \newline and MSR-86K subset  & 10.79\% \\ 
        Qwen2.5-7B-Instruct & competition speech \newline and MSR-86K subset & 10.01\% \\ \midrule
        Whisper Turbo  & competition speech & 12.97\% \\
        Whisper-large-v3  & competition speech & 11.61\% \\
        Whisper-large-v3 & competition speech \newline and MSR-86K subset & 10.4\% \\   
    \bottomrule
    \end{tabular}
\end{table}

\begin{figure}[t]
  \centering
  \includegraphics[width=\linewidth]{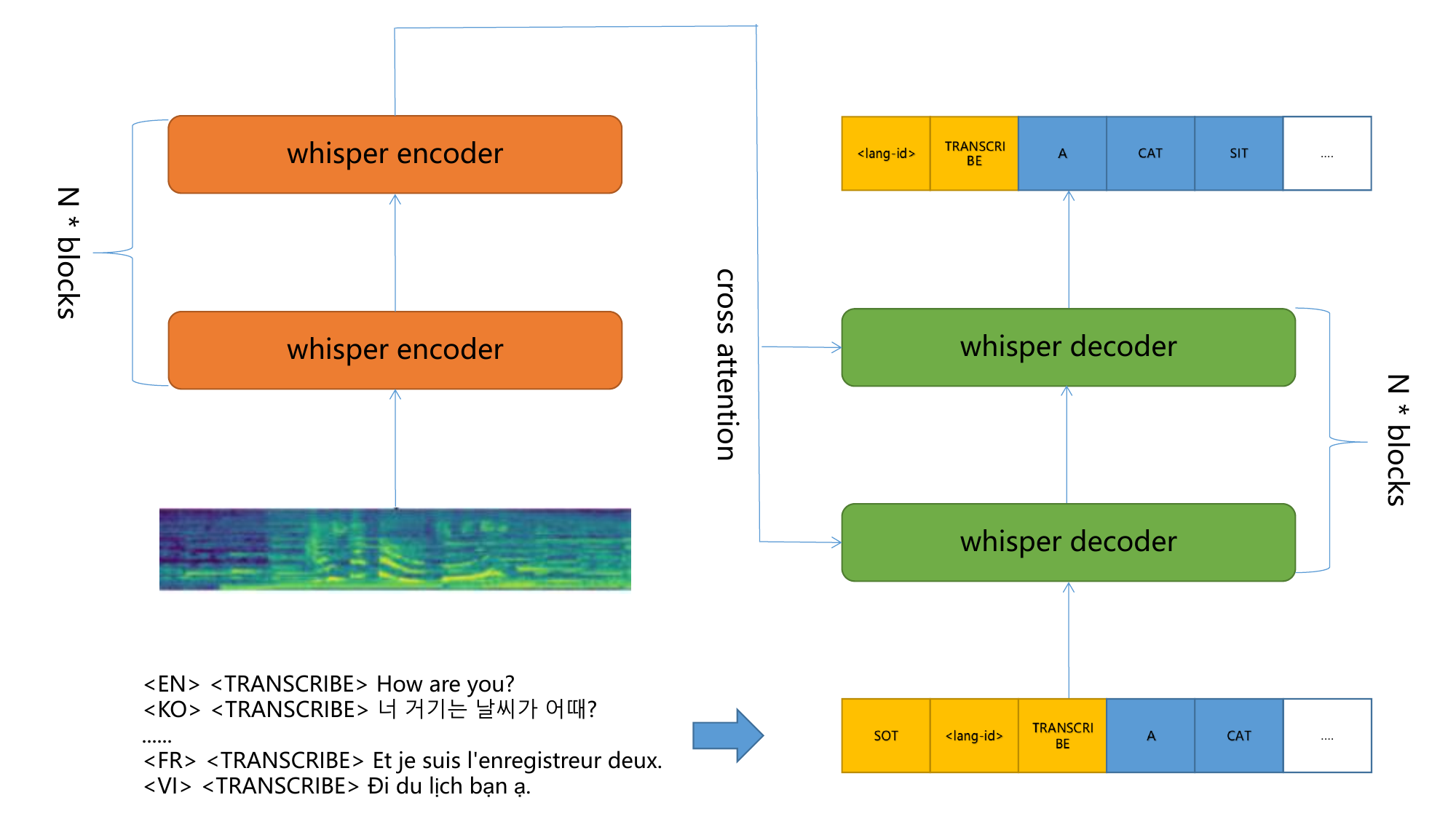}
  \caption{Whisper Multilingual Finetuning System Architecture.}
  \label{fig:speech_production}
\end{figure}

\subsection{Final Submission system}

Building upon the insights gained from the ablation studies, the development(Dev) set was incorporated into the final training dataset, and the model was further fine-tuned for an additional 2 epochs. Our final submission system achieved a word/character error rate (WER/CER) of 9.83\% across 11 languages in the evaluation set, showcasing its effectiveness and performance.

\section{Conclusions}
This paper presents a Multilingual Automatic Speech Recognition (ASR) framework, developed for the MLC-SLM 2025 Challenge, based on an Encoder-Adaptor-LLM architecture. The system incorporates a frozen Whisper-large-v3 encoder for robust acoustic feature extraction, a trainable adaptor to ensure efficient alignment between speech and text representations, and a frozen Qwen2.5-7B-Instruct LLM fine-tuned using Low-Rank Adaptation (LoRA) for task-specific optimization. Through a meticulously designed training pipeline and ablation studies, the effectiveness of data augmentation and model scaling in improving performance was thoroughly demonstrated. This work highlights the crucial role of integrating advanced architectures with diverse and high-quality datasets to advance multilingual ASR capabilities.

\bibliographystyle{IEEEtran}
\nocite{*}  

\bibliography{mybib}

\end{document}